\documentclass[12pt]{article}
\usepackage{epsfig,amsmath,amssymb,here,rotating}
\usepackage[T1]{fontenc}
\usepackage{eurosym}

\def\dm{\ensuremath{\Delta m}}
\def\dm2{\ensuremath{\Delta m^{2}\ }}
\def\sen2th{\ensuremath{ \sin^{2}(2\theta)\ }}
\def\(({\left(}
\def\)){\right)}

\def\nue{\ensuremath{\nu_{e}\ }}
\def\nubare{\ensuremath{\overline{\nu}_{e}\ }}

\def\numu{\ensuremath{\nu_{\mu}\ }}
\def\nubarmu{\ensuremath{\overline{\nu}_{\mu}\ }}

\newcommand{\be}{\begin{equation}}
\newcommand{\ee}{\end{equation}}

\newcommand{\dmtt}{\ensuremath{\Delta m^2_{23} \,}}

\newcommand{\He}{\ensuremath{^6{\mathrm{He}\,}}}
\newcommand{\Ne}{\ensuremath{^{18}{\mathrm{Ne}\,}}}

\newcommand{\thetaot}{\ensuremath{\theta_{13}}\,}

\newcommand{\sigdm}{\ensuremath{{\rm sign}(\Delta m^2_{23})\ }}

\def\He{\ensuremath{^6{\mathrm{He}}}}

\def\Ne{\ensuremath{^{18}{\mathrm{Ne}}}}

\def\mc2{\multicolumn{2}{c|}}
\def\beq{\begin{equation}}
\def\eeq{\end{equation}}
\def\bea{\begin{eqnarray}}
\def\eea{\end{eqnarray}}
\def\bq{\begin{quote}}
\def\eq{\end{quote}}

\def\gappeq{\mathrel{\rlap {\raise.5ex\hbox{$>$}}{\lower.5ex\hbox{$\sim$}}}}
\def\lappeq{\mathrel{\rlap{\raise.5ex\hbox{$<$}}{\lower.5ex\hbox{$\sim$}}}}

\begin{document}

\bibliographystyle{JHEP}

\begin{titlepage}


\begin{center}
{\bf \Large MEMPHYS\,:}

\vspace{0.2cm} 

{\bf \Large A large scale water \v{C}erenkov detector at Fr\'ejus}


\end{center}

\vspace{2.5cm}

\begin{center}

{A. de Bellefon$^{(1)}$, 
J. Bouchez$^{(1)}$$^{(2)}$, 
J. Busto$^{(3)}$, 
J.-E. Campagne$^{(4)}$, \\
C. Cavata$^{(2)}$, 
J. Dolbeau$^{(1)}$,
J. Dumarchez$^{(5)}$, 
P. Gorodetzky$^{(1)}$,
S. Katsanevas$^{(1)}$, \\
M. Mezzetto$^{(6)}$,  
L. Mosca$^{(2)}$, 
T. Patzak$^{(1)}$, 
P. Salin$^{(1)}$,
A. Tonazzo$^{(1)}$, 
C. Volpe$^{(7)}$}

\vspace{0.5cm}
{\it $^{(1)}$ APC Paris \\
     $^{(2)}$ DAPNIA-CEA Saclay \\
     $^{(3)}$ CPP Marseille \\
     $^{(4)}$ LAL Orsay \\
     $^{(5)}$ LPNHE Paris \\
     $^{(6)}$ INFN Padova \\ 
     $^{(7)}$ IPN Orsay
}

\end{center}
\vspace{1.5cm}
\begin{center}
{\bf Abstract}\\
A water \v{C}erenkov detector project, of megaton scale, to be installed
in the Fr\'ejus underground site and dedicated to nucleon decay,
neutrinos from supernovae, solar and atmospheric neutrinos, as well as
neutrinos from a super-beam and/or a beta-beam coming from CERN, is
presented and compared with competitor projects in Japan and in the
USA. The performances of the European project are discussed, including
the possibility to measure the mixing angle $\theta_{13}$ and the
CP-violating phase $\delta$. 

\end{center}

\vspace{2.3cm}
\end{titlepage}

\newpage
\tableofcontents
\newpage

\section{Motivations}

There is a steady 25 year long tradition of 
water \v{C}erenkov observatories having produced an  
incredibly rich harvest of seminal discoveries. 
The water \v{C}erenkov movement was started in the early 80's 
by the scientists searching for proton decay. 
It fulfilled indeed this purpose by extending the proton decay 
lifetimes a few orders of magnitude.  Furthermore, water \v{C}erenkov's, 
through a serendipitous turn, as frequently happens in physics, 
have also inaugurated:  
\begin{itemize}
\item particle astrophysics through the detection of the neutrinos 
coming from the explosion of the supernova 1987a  
by IMB and Kamioka, acknowledged by the Nobel prize for Koshiba 
\item the  golden era of neutrino mass and oscillations by discovering 
hints for atmospheric neutrino oscillations while at the same time 
confirming earlier solar neutrino oscillation results. 
\end{itemize}
The latest in the water \v{C}erenkov series, 
the well known Super-Kamiokande, has now given strong evidence 
for a maximal oscillation between
\numu and $\nu_\tau$, 
and several projects with accelerators have
been designed 
to check this result. The results of the K2K
experiment confirm the oscillation, and other experiments (MINOS in the USA,
OPERA and ICARUS at Gran Sasso) should refine most of 
the oscillation parameters by 2010.

More recently, after the results from SNO and KamLAND, a solid
proof for solar neutrino flavour oscillations governed by the
so-called LMA solution has been established. We can no longer
escape the fact that neutrinos have indeed a mass, although the
absolute scale is not yet known. Furthermore, the large mixing
angles of the two above-mentioned oscillations and their relative
frequencies open the possibility to test CP violation in the
neutrino sector if the third mixing angle, $\theta_{13}$, is not
vanishingly small (we presently have only an upper limit at about
0.2 on $\sin^2(2\theta_{13})$, provided by the CHOOZ experiment). Such a
violation could have far reaching consequences, since it is
a crucial ingredient of leptogenesis, one of the
presently preferred explanations for the matter dominance in our
Universe.

The ideal tool for these studies is thought to be the so-called neutrino
factory, which would produce through muon decay intense neutrino beams
aimed at magnetic detectors placed several thousand kilometers away from
the neutrino source.

However, such projects would probably not be launched unless
one is sure that the mixing angle $\theta_{13}$, governing the oscillation
between \numu and \nue at the higher frequency, is such that this oscillation
is indeed observable.
This is why physicists have considered the possibility
of producing new conventional neutrino beams of unprecedented intensity, made
possible by recent progress on the conception of proton drivers with a factor
10 increase in power (4 MW compared to the present 0.4 MW of the FNAL beam).
While the present limit on $\sin^2(2\theta_{13})$ is around 0.2, 
these new neutrino
``superbeams'' would explore $\sin^2(2\theta_{13})$ down to 
$2\cdot 10^{-3}$ (i.e a factor
100 improvement on the \numu - \nue  oscillation amplitude).

European working groups have studied a neutrino factory at CERN
for some years, based on a new proton driver of 4 MW, the SPL.
Along the lines described above, a subgroup on neutrino
oscillations has studied the potentialities of a neutrino
superbeam produced by the SPL. The energy of produced neutrinos is
around 300 MeV, so that the ideal distance to study \numu to \nue
oscillations happens to be 130 km, that is exactly the distance
between CERN and the existing Fr{\'e}jus laboratory. The present
laboratory cannot house a detector of the size needed to study
neutrino oscillations, which is around 1 million cubic meters. But
the recent decision to dig a second gallery, parallel to the
present tunnel, offers a unique opportunity to realize the needed
extension for a reasonable price.

Due to the schedule of the new gallery, a European project would
be competitive only if the detector at Fr{\'e}jus reaches a
sensitivity on $\sin^2(2\theta_{13})$ around $10^{-3}$, since other projects in
Japan (T2K phase 1) and USA (NoVA) will have reached $10^{-2}$
by 2015. The working group has then decided to study
directly a water \v{C}erenkov detector with a mass approaching 1 megaton,
necessary to reach the needed sensitivity. This detector has been
nicknamed MEMPHYS (for MEgaton Mass PHYSics). Its study has benefited from a
similar study by our American colleagues, the so-called UNO
detector with a total mass of 660 kilotons. Simulations have shown
that the sensitivity on $\sin^2(2\theta_{13})$ at a level of $10^{-3}$  
could indeed be fulfilled with MEMPHYS.

This version of the project 
as two competitors,
since japanese and
american physicists have their own project, with similar
potentialities. But owing to a new idea recently proposed
by Piero Zucchelli, the european project could have a unique
characteristics which would make it very appealing. This idea
is to send towards Fr{\'e}jus, together with the SPL superbeam,
another kind of neutrino beam, called beta beam, made of \nue or \nubare 
produced by radioactive nuclei stored in an accumulation ring.
CERN has a very good expertise on the production and acceleration
of radioactive nuclei. Studies show that such beams would reach
performances even better than those of the SPL on the oscillation
between \nue and \numu, with a sensitivity on $\theta_{13}$ down to half a
degree, with a factor four gain. 
But the main point is that both beams, if
run simultaneously, would allow to study the violation of CP
symmetry in a much more efficient and redundant
way than when using only the SPL
beam. This peculiarity, which would be a CERN exclusivity, would
give a considerable bonus to our project concerning neutrino
studies, since it could reach sensitivities on CP violation as
good as those of a neutrino factory for $\sin^2(2\theta_{13})$ above 
$5\cdot 10^{-3}$.

As mentioned in the beginning, 
such a detector will not only do the physics of neutrino oscillations, 
but would also address equally fundamental questions in 
particle physics and particle astrophysics. 

In particular, such a detector could
reach a sensitivity around $10^{35}$ years on the proton lifetime, which is
precisely the scale at which such decays are predicted by most supersymetric
or higher dimension grand unified theories, thus giving the hope 
for a fundamental discovery.

Such a detector would also bring a wealth of information on supernova
explosions: it would detect more than $10^5$ neutrino interactions
within a few seconds if such an explosion occurs in our galaxy, and 
would observe a statisticaly significant signal for explosions at distances 
up to 1 Mpc, and provide a supernova trigger to other astroparticle detectors
(gravitational antennas and neutrino telescopes). 
For galactic supernova explosions, the huge available 
statistics would give access to a detailed description of 
the collapse mechanism 
and neutrino oscillation parameters. In addition, the huge mass of the detector
could allow to detect for the first time
the diffuse neutrinos from past SN explosions.

The proposed detector is indeed a multipurpose detector addressing several
issues of utmost importance.

\newpage

\section{Megaton Physics}
\subsection{Proton decay}

Proton decay is one of the few predictions of Grand Unified Theories
that can be tested in low-energy experiments. Its discovery would
definitely testify for a more fundamental structure beyond the Standard
Model.

In the past twenty years, the first generation (IMB, Fr{\'e}jus, Kamiokande)
and second generation (Super-Kamiokande) proton decay experiments have
already put stringent lower limits on the partial proton lifetimes,
qualitatively ruling out non-supersymmetric $SU(5)$ theories (first generation)
and the minimal supersymmetric $SU(5)$ theory (second generation). 
A megaton-scale
water \v{C}erenkov detector would improve further the experimental sensitivity
to proton decay by more than one order of magnitude and allow to probe
non-minimal $SU(5)$ models as well as other types of GUTs, such as $SO(10)$,
flipped $SU(5)$ and higher-dimensional GUTs.
Indeed, recent experimental and theoretical progresses point
towards smaller values of the partial lifetime of the proton into
$\pi^0 e^+$, implying that this decay mode -- the most
model-independent one -- is not out of reach, contrary to previous
expectations. Using the new, more accurate lattice calculation of
the nucleon decay matrix element one can estimate 
$\tau (p \rightarrow \pi^0 e^+) 
\approx 10^{35}\, \mbox{yrs}\, (M_X / 10^{16}\, 
\mbox{GeV})^4\,
((1/25) / \alpha_{GUT})^2$, 
where $M_X$ is the mass of the
superheavy gauge bosons mediating proton decay, 
$\alpha_{GUT}
\equiv g^2_{GUT} / 4 \pi$ and $g_{GUT}$ is the value of the GUT
gauge coupling at the unification scale. This is to be compared
with the present Super-Kamiokande lower limit ($5 \times 10^{33}\,
\mbox{yrs}$), and with the expected sensivity of a megaton water
\v{C}erenkov detector ($10^{35}\, \mbox{yrs}$ after 10 years 
of data taking for MEMPHYS). 

The dominant decay channel in supersymmetric GUTs,
$p \rightarrow K^+ \bar \nu$, is much more model-dependent. The corresponding
decay rate indeed depends on the couplings and masses of the supersymmetric
partners of the heavy colour-triplet Higgs bosons, and on the details of the
sparticle spectrum. The effective triplet mass, in particular, is extremely
dependent on the GUT model.
In many models, one finds an upper limit $\tau (p \rightarrow K^+ \bar \nu)
\leq \mbox{few}\, 10^{34}\, \mbox{yrs}$ 
\cite{Dermisek:2000hr}\cite{Babu:1998wi}\cite{Altarelli:2000fu},
to be compared with the present Super-Kamioka nde lower limit
($1.6 \times 10^{33}\, \mbox{yrs}$), and with the expected sensivity of
a megaton water \v{C}erenkov detector ($2 \times 10^{34}\, \mbox{yrs}$ after
10 years for MEMPHYS). 

There are many more decay channels that could be accessible to a megaton
water \v{C}erenkov detector. The measurement of several partial lifetimes
would allow to discriminate between different Grand Unified models, at a time
when, after several years of LHC running, 
the supersymmetry landscape will be drastically clarified, 
through discovery or severe exclusion limits. Therefore the predictions 
of proton lifetime, in constrained or more general supersymmetric models, 
will be sharpened even further.

\subsection {Supernovae}

The core collapse supernovae are spectacular events which have been
theoretically studied for more than three decades. After explosion the star
loses energy, mainly by neutrino emission, and cools down, ending as a
neutron star or a black hole. Many features of the collapse mechanism are
indeed imprinted in the neutrinos released during the explosion.

At the same time, a galactic supernova would give particle physicists the
occasion to explore the neutrino properties on scales of distance up to
$10^{17}$ km and of time up to $\sim 10^{5}$ years and at very high density.
The detected signal from a supernova explosion depends on the
structure of the neutrino mass spectrum and lepton mixing. Therefore, in
principle, studying the properties of a supernova neutrino burst one
can get information about the values of parameters relevant for the
solution of the solar neutrino problem, the type of the mass
ordering (the so-called mass hierarchy), 
the mixing parameter $\sin^2\theta_{13}$, the
presence of sterile neutrinos and new neutrino interactions.

It is generally believed that
core-collapse supernovae have occurred throughout the
Universe since the formation of stars. Thus, there should
exist a diffuse background of neutrinos originating from
all the supernovae that have ever occurred. Detection
of these diffuse supernova neutrinos (DSN) would offer 
insight about the history of star formation and supernovae
explosions in the Universe.

Now the requirements for a detector are to be very massive, located
underground, to stay in operation for at least 20 years and to be equipped
with a real time neutrino detection electronics with a threshold around
10 MeV. For those reasons a megaton water \v{C}erenkov detector with a
fiducial volume around 450 kt is a good choice. Such a detector would
detect $\sim 10^{5}$
events from a galactic stellar collapse, and of the order of 20 events from a
supernova in Andromeda galaxy, which is one of the closest to our Milky way.
The large mass of such a detector compared to other proposed and existing
facilities means that the sample collected will outnumber that of all other
detectors combined.
The general and relative performances are summarized in section \ref{sec:SN}.


All types of neutrinos and anti-neutrinos are emitted
from a core-collapse supernova, but not all are equally
detectable. 
The $\bar{\nu_e}$ is most likely to interact in a water \v{C}erenkov detector.
Three main neutrino signals would be detected, each one yielding unique
information: 
\begin{enumerate}
\item Inverse beta decay events (89\%) allowing for a good
determination of the time evolution and energy distribution of the neutrino
burst. The potentials would be enhanced by the detection of the neutron
with the addition of a small mount of Gadolinium \cite{Beacom:2003nk}.
\item Neutral current events involving $^{16}O$ (8\%), which are sensitive
to the temperature of the neutrino spectrum.
\item Directional elastic scattering events from $\nu_{x}$ + $e^{-}$ and
$\bar{\nu}_{x}$ + $e^{-}$ ($\sim$ 3\%). These events provide the direction
of the supernova within $\pm 1$ degree.
\end{enumerate}

\subsection {$\theta_{13}$ and CP violation in oscillations}

In the recent years, a series of experiments have provided strong evidence
for oscillations of solar and atmospheric neutrinos, and have started to
precisely constrain the associated parameters $\Delta m^2_{23}$,
$\Delta m^2_{12}$, $\theta_{23}$ and $\theta_{12}$. The third mixing angle
$\theta_{13}$ is still unknown: all we have is an upper bound of
$\theta_{13} \leq 13^\circ$ coming from the CHOOZ 
experiment \cite{Apollonio:1999ae}. Its
measurement, as well as the determination of the sign of $\Delta m^2_{23}$
and therefore of the type of mass hierarchy, is crucial for
discriminating between different neutrino mass and mixing scenarios.
Moreover a precise determination of the PMNS matrix (which contrary
to the CKM matrix is free from hadronic uncertainties) would put very severe
constraints on models
of fermion masses, including realistic GUT models, and thus shed some light
on the underlying flavour theory. A neutrino super-beam from the CERN
SPL to a megaton water \v{C}erenkov detector located at Fr\'ejus would allow
to make a significant progress in this programme, reaching in particular
a sensitivity to $\sin^2(2\theta_{13})$ close to $10^{-3}$ 
and close to $2\cdot 10^{-4}$ 
with a Beta-beam (and $1\cdot 10^{-4}$ 
with both Super-beam and
Beta-beam), see section~\ref{sec:oscillations}.

Due to its sensitivity to $\theta_{13}$, a megaton water \v{C}erenkov
detector would also be sensitive to the CP violating phase
$\delta$ in a large portion of the ($\Delta m^2_{12}$,
$\theta_{13}$) parameter space. Establishing CP violation in the
lepton sector would represent a major progress in particle
physics, since CP violation has only been observed in the quark
sector so far. Moreover, CP violation is a crucial ingredient of
leptogenesis, a mechanism for creating the matter-antimatter
asymmetry of the Universe which relies on the out-of-equilibrium
decay of heavy Majorana neutrinos. Although the phase involved in
oscillations is generally distinct from the phase responsible for
leptogenesis, the measurement of a nonzero $\delta$ would be a
strong indication that leptogenesis may be at the origin of the
baryon asymmetry \cite{Fukugita:1986hr}. 
Indeed, standard electroweak baryogenesis would
require a very light Higgs boson, which is now excluded by LEP,
and only a small window remains for supersymmetric electroweak
baryogenesis.
Another necessary ingredient of leptogenesis is the existence of Majorana
neutrinos, which could be established by a positive signal in future
neutrinoless double beta decay experiments.

\newpage
\section{Underground laboratory and detector} 

\subsection{Results of a feasibility study in the 
central region of the Fr\'ejus tunnels}
\label{sec:undlab}
\begin{figure}[htb]
\centerline{\epsfig{figure=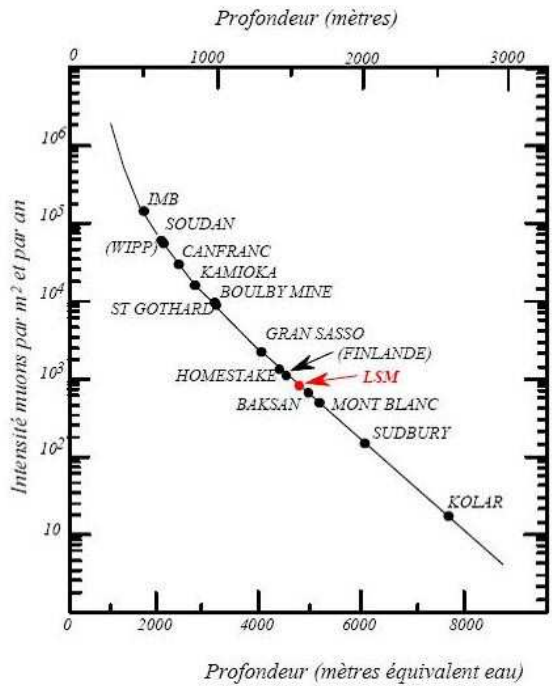,width=8cm}}
\caption{\it Muon flux as a function of overburden. 
The Frejus site is indicated by "LSM".}
\label{muonflux}
\end{figure}

The site located in the Fr\'ejus mountain in the Alps, 
which is crossed by a road-tunnel connecting France (Modane) 
to Italy (Bardonecchia), has a number of interesting characteristics 
making it a very good candidate for the installation of a megaton-scale 
detector in Europe, aimed both at non-accelerator and accelerator 
based physics.
Its great depth (4800 mwe, see figure~\ref{muonflux}), 
the good quality of the rock, 
the fact that it offers horizontal access, its distance from CERN (130 km), 
the opportunity of the excavation of a second (``safety'') tunnel, 
the very easy access by train (TGV), by car (highways) 
and by plane (Geneva, Torino and Lyon airports),
the strong support from the local authorities
represent the most important of these characteristics.

On the basis of these arguments, the DSM (CEA) and IN2P3 (CNRS) institutions 
decided to perform a feasibility study of a Large Underground Laboratory 
in the central region of the Fr\'ejus tunnel, near the already existing, 
but much smaller, LSM Laboratory. This preliminary study
has been performed by the SETEC (French) and STONE (Italian) companies 
and is now completed. These companies already made the study and managed 
the realisation of the Fr\'ejus road tunnel and of the LSM 
(Laboratoire Souterain de Modane) Laboratory.
A large number of precise and systematic measurements of the rock 
characteristics, performed at that time, have been used to make a 
pre-selection of the most favourable regions along the road tunnel 
and to constrain the simulations of the present pre-study for the 
Large Laboratory. 

\begin{figure}[htb]
\centerline{\epsfig{figure=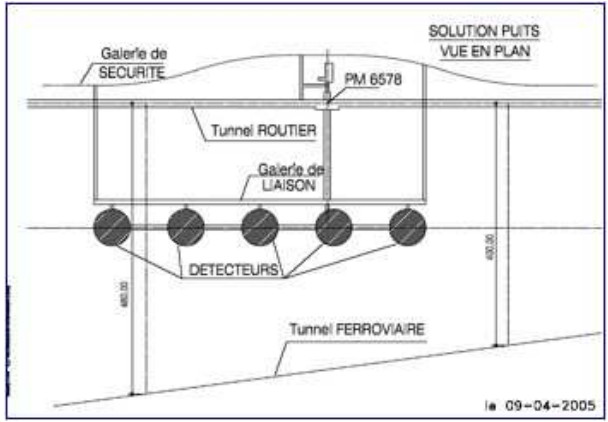,width=8cm,}}
\caption{\it Possible layout of the Fr\'ejus underground laboratory.}
\label{fig:tunnel}
\end{figure}

Three regions have been pre-selected : 
the central region and two other regions at about 3 km from 
each entrance of the tunnel. 
Two different shapes have been considered for the cavities to be excavated: 
the ``tunnel shape'' and the cylindrical ``shaft shape''. The main purpose 
was to determine the maximum possible size for each of them, 
the most sensitive dimension being the width (the so-called ``span'') 
of the cavities. 	

\begin{figure}[htb]
\centerline{\begin{tabular}{cc}
\epsfig{figure=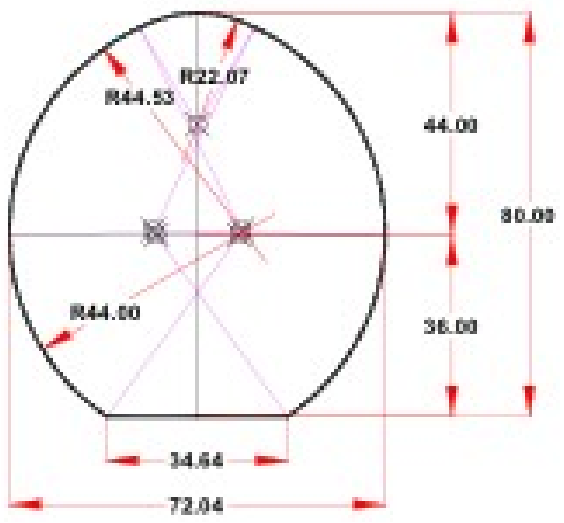,width=8.5cm} &
\epsfig{figure=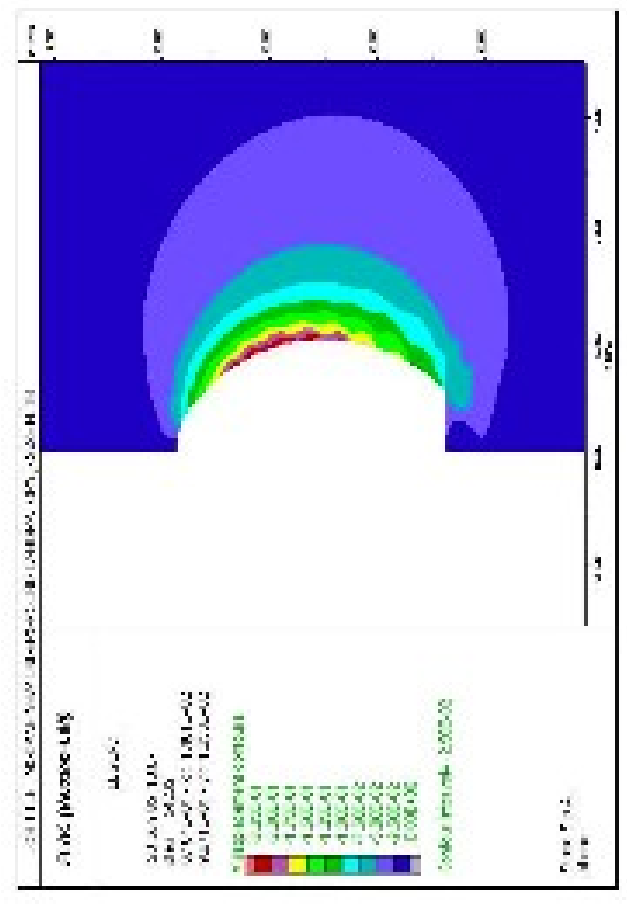,width=5.5cm} 
\end{tabular}}
\caption{\it An example of ``egg shape'' simulation, 
constrained by the rock parameter measurements made during the road tunnel 
and the present laboratory excavation.
The main feasibility criterium is that the significantly perturbated 
region around the cavity should not exceed a thickness of about 10 m.}
\label{fig:egg}
\end{figure}

The very interesting results of this preliminary study 
can be summarized as follows : 
\begin{enumerate}
\item the best site (rock quality) is found in the middle of the mountain, 
at a depth of 4800 mwe;
\item of the two considered shapes : ``tunnel'' and ``shaft'', 
the ``shaft shape'' is strongly preferred;
\item cylindrical shafts are feasible up to a diameter $\Phi$ = 65 m  
and a full height  h = 80 m ($\sim$ 250000 m$^3$);
\item with ``egg shape'' or ``intermediate shape between 
cylinder and egg shapes'' the volume of the shafts could be still increased
(see Fig.~\ref{fig:egg});
\item the estimated cost is $\sim$ 80 M Euro per shaft.
\end{enumerate}

Fig.~\ref{fig:tunnel} shows a possible configuration for this
large Laboratory, where up to five shafts, of about 250000 m$^3$ each, 
can be located between the road tunnel and the railway tunnel, 
in the central region of the Fr\'ejus mountain. 

Two possible scenarios for Water \v{C}erenkov detectors are, for instance: 
\begin{itemize}
\item 3 shafts of 250000 m$^3$ each, with a fiducial mass of 440 kton 
(``UNO-like'' scenario).
\item 4 shafts of 250000 m$^3$ each, with a fiducial mass of 580 kton.
\end{itemize}
In both scenarios one additional shaft could be excavated
for a Liquid Argon and/or a liquid scintillator detector of about 
100 kton total mass.

The next step will be a Design Study for this Large Laboratory, 
performed in close connection with the Design Study of the detectors 
and considering the excavation of 3 to 5 ``shafts'' of about 250 000 m$^3$ 
each, the associated equipments and the mechanics of the detector modules. 

\subsection{Detector: general considerations}

The 20 year long successful operation of the 
Super-Kamiokande detector has clearly
demonstrated the capabilities and limitations of large water \v{C}erenkov
detectors\,:
\begin{itemize}
\item This technique is by far the cheapest 
and the most stable to instrument a very large detector
mass, as price is dominated by the photodetectors and their associated
electronics (this price growing like the outer surface of the detector),
while the active mass, made of water, is essentially free except for the 
purification system
\item These detectors are mainly limited in size by the finite attenuation
length of \v{C}erenkov light, found to be 80 meters at $\lambda = 400$~nm
in Super-Kamiokande, and by the pressure of water on the photomultipliers at
the bottom of the tank, which gives a practical limit of 80~m in height.
At large depths, the maximal size of underground cavities actually limits
relevant dimensions to about 70~m.
\item The detection principle consists in measuring \v{C}erenkov rings produced
by charged particles going faster than light in water. This has several
consequences\,:
\begin{enumerate}
\item neutral particles and charged particles
below \v{C}erenkov threshold are undetectable, so that some energy may
be missing
\item complicated topologies are difficult to handle, and in practice only
events with less than 3 to 5 rings are efficiently reconstructed
\item ring topology, based on their degree of fuzziness, allows to separate
between electromagnetic (e, $\gamma$) rings and ($\mu$, $\pi$) rings
\item the threshold in particle energy depends mainly on photocathode
coverage and also on water purity (due to radioactive backgrounds, such as
radon). Super-Kamiokande has achieved an energy threshold of 5 MeV with
40\% cathode coverage
\item due to points 1 and 2, water \v{C}erenkov detectors are not suited to
measure high energy neutrino interactions, as more rings and more undetectable
particles are produced. A further limitation comes from the confusion between
single electron or gamma rings and high energy $\pi^0$'s giving 2 overlapping
rings. In practice water \v{C}erenkov's stay excellent neutrino detectors
for energies below 1 (may be 2) GeV, when interactions are mostly quasi-elastic 
and the 2 rings from $\pi^0$ well separated.
\end{enumerate}
\end{itemize}
  
\subsection{Detector design}

Three detector designs are being carried out worldwide,
namely Hyper-Kamiokande \cite{uno} in Japan,
UNO \cite{hyperk} in the USA  and the present project MEMPHYS in Europe.
All of them are rather mild extrapolations of Super-Kamiokande, and rely on the 
expertise acquired after 20 years of operation of this detector.
Their main characteristics are summarized in table~\ref{WC:tab-1}.
\begin{sidewaystable}
\centering
\begin{tabular}{rccc}
\hline\noalign{\smallskip}
 Parameters                  &        \textbf{UNO} (USA)            &   
\textbf{HyperK} (Japan)          &      \textbf{MEMPHYS} (Europe)\\
\noalign{\smallskip}\hline\noalign{\smallskip}
\multicolumn{4}{l}{\textbf{Underground laboratory}}  \\ 
       location   &   Henderson / Homestake      &   Tochibora    & Fr\'ejus \\
                depth (m.e.w.)  &    4500/4800                 &    
1500                  &        4800  \\
Long Base Line (km)   & $1480\div2760$ / $1280\div2530$ & 290    &   130 \\
                       & FermiLab$\div$BNL       & JAERI         &   CERN \\
\noalign{\smallskip}\hline\noalign{\smallskip}
\multicolumn{4}{l}{\textbf{Detector dimensions}}          \\
type              & 3 cubic compartments   & 2 twin tunnels  & $3\div5$
shafts\\
                  &                        & 5 compartments  &           \\
dimensions            & $3\times (60\times60\times60)\mathrm{m}^3$ 
                                                                        &
$2\times 5 \times (\phi=43\mathrm{m} \times L=50\mathrm{m})$       &
$(3\div5)\times(\phi=65\mathrm{m} \times H=65\mathrm{m}) $ \\   
fiducial mass (kt)& 440                          &       550                  
& $440\div730$\\
\noalign{\smallskip}\hline\noalign{\smallskip}
\multicolumn{4}{l}{\textbf{Photodetectors}}          \\
           type   & 20" PMT              & 13" H(A)PD           & 12" PMT
\\
    number (internal detector)     & 57,000      & 20,000 per compartment
                                                & 81,000 per shaft \\
 surface coverage & 40\% (1/3) \& 10\% (2/3) & 40\%  & 30\%\
                                            
\\
\noalign{\smallskip}\hline
\end{tabular}
\caption{\label{WC:tab-1}
\it Some basic parameters of the three Water \v{C}erenkov
detector baseline designs}
\end{sidewaystable}

These 3 projects aim at a fiducial mass around half a megaton, taking into
account the necessity to have a veto volume on the edge of the detector, 1 to 2
meters thick, plus a minimal distance of about 2 meters between photodetectors
and interaction vertices, leaving some space for ring development.
The main differences between the 3 projects lie in the geometry 
of the cavities (tunnel shape for Hyper-Kamiokande, shafts for MEMPHYS,
intermediate with 3 cubic modules for UNO), and the photocathode coverage,
similar to Super-Kamiokande for Hyper-Kamiokande 
and MEMPHYS, while UNO keeps this
coverage on only 1 cubic detector, while the 2 others have only 10\% coverage
for cost reasons. Another important parameter is the rock overburden, similar
for UNO and MEMPHYS (4800~mwe), but smaller for Hyper-Kamiokande (1500~mwe),
which might be a limiting factor for low energy physics, due to spallation
products and fast neutrons produced by cosmic muons, more
abundant by 2 orders of magnitude (see figure~\ref{muonflux}).

The basic unit for MEMPHYS consists of a cylindrical detector module 65~meters
in diameter and 65~meters high, which can be housed
in a cylindrical cavity with 70~meter diameter and 80 meter height, as proven
by the prestudy. This corresponds to a water mass of 215 kilotons, that is
only 4 times the Super-Kamiokande detector. 
Conservatively substracting 2~m for the
outer veto plus 2~m for the fiducial volume, this leaves us with a fiducial
mass of 146 kilotons per module. The baseline design uses 3 modules, giving
a total fiducial mass of 440 kilotons, like UNO, corresponding to factor
20 increase over Super-Kamiokande (4 modules would give 580 
kiloton fiducial mass). The modular aspect is actually mandatory for
maintenance reasons, so that at least 2 of the 3 modules would be active
at any time, giving 100\% duty cycle for supernova explosions. Furthermore,
it would offer the possibility to add Gadolinium in one of the modules, which
has been advocated to improve diffuse supernova neutrino detection.
We estimate an overall construction time of less than 10 years, and of course
the first module could start physics during the completion of the two other
modules.

\subsection{Photodetection}

The baseline photodetector choice is photomultipliers (PMT) as they have
successfuly equipped the previous generation of large water \v{C}erenkov
detectors and many other types of presently running detectors in HEP. The PMT
density should be chosen to allow excellent sensitivity to a broad range
of nucleon decays and neutrino physics while keeping the instrumentation costs
under control.

Our goal for MEMPHYS is to reach in the whole detector
the same energy threshold as Super-Kamiokande,
that is 5 MeV, important for solar neutrino studies, for the proton decay into
$K^+ \nu$ using the 6~MeV tag from $^{15}$N desexcitation, and also very useful
for SN explosions, since the measurement of the $\nu_{\mu}$ and $\nu_{\tau}$
fluxes could be achieved using the neutral current excitation of Oxygen.

Our first approach was to consider 20" Hamamatsu tubes as used by
Super-Kamiokande, but the cost for 40\% coverage becomes prohibitive, as these
tubes are manually blown by specially trained people, which makes them very
expensive. Following a suggestion presented at the NNN05 conference by Photonis
company, we have considered the possibility of using instead 12" PMT's, which
can be automatically manufactured and have better characteristics
compared to 20" tubes\,: quantum efficiency (24\% vs 20\%), collection
efficiency (70\% vs 60\%), risetime (5~ns vs 10~ns), jitter (2.4~ns vs 5.5~ns).
Based on these numbers, 30\% coverage with 12" PMT's would give the same number
of photoelectrons per MeV as a 40\% coverage with 20" tubes. Taking into 
account the ratio of photocathodes (615~cm$^2$ vs 1660~cm$^2$), this implies
that going from 20" tubes to twice as many 12" tubes will give the same
detected light, with a bonus on time resolution and on pixel locations. If the dark
current of better photocathodes does not increase dramatically the trigger rate, 
we can expect MEMPHYS performances be at least as good as Super-Kamiokande.
A GEANT4 based Monte Carlo is under development to quantify the effective
gain. Pricewise, each 20" PMT costing 2500 Euros 
is replaced by 2 12" PMT's costing 800 Euros each. 
The only caveat is to make sure that the savings on PMT's are not
cancelled by the doubling of electronic channels. An R\&D on electronics
integration is presently underway (see Sec. \ref{sec:photo}).

\subsection{Photomultiplier tests}

A joint R\&D program between Photonis company and French laboratories has been
launched to test the quality of the 12" PMTs in the foreseen conditions of 
deep water depth, and to make a realistic
market model for the production of about 250,000 PMTs that would be necessary
to get the 30\% geometrical coverage. 

In parallel, studies on new photo-sensors have been launched. 
The aim is to reduce cost, while improving production rate and performance, as
it is essential
to achieve the long term stability and reliability which is proven for PMTs. 
Hybrid photosensors (HPD) could be a solution\,:
the principle has been proven by ICRR and Hamamatsu with a 5" HPD prototype. 
Successful results from tests of an 13"
prototype operated with 12 kV are now available, 
showing a $3 \cdot 10^4$ gain, good single photon sensitivity, 
0.8 ns time resolution and a satisfactory gain and 
timing uniformity over the photo-cathode area.
The development of HPD has also been initiated in Europe, 
in collaboration with Photonis.

\subsection{Smart-photodetector electronics}
\label{sec:photo}
\begin{figure}[htb]
 \begin{minipage}[c]{0.44\textwidth}
\centering
\includegraphics[width=0.85\textwidth]{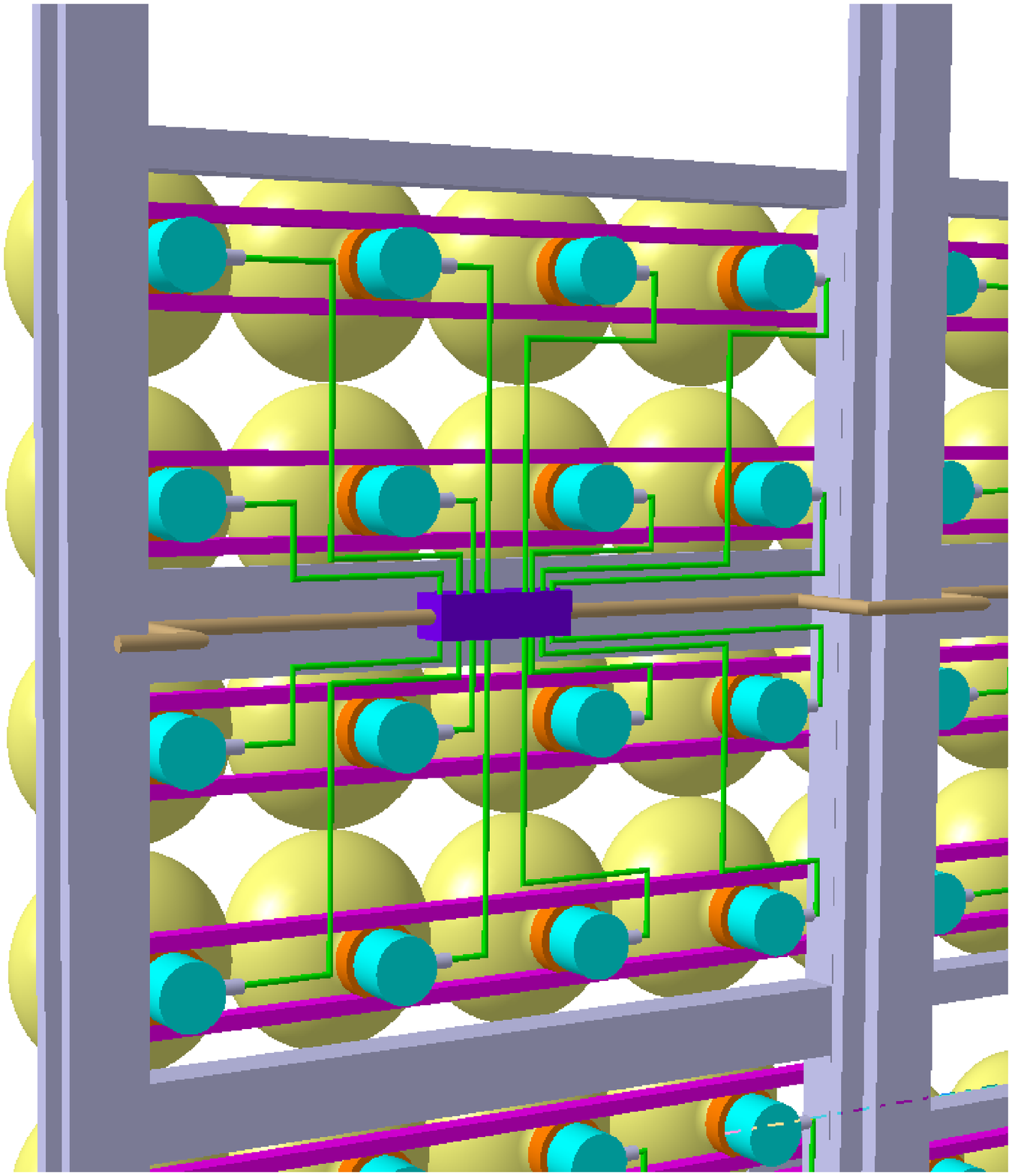}
\caption{\label{fig:MEMPHYSPMTS}
\it Sketch of a possible photosensor basic module composed of a matrix
of $4\times4$ 12" PMTs with the electronic box containing the High
Voltage unit and the Readout chip.}	 
\end{minipage}
 \begin{minipage}[c]{0.04\textwidth}
~
\end{minipage}
 \begin{minipage}[c]{0.44\textwidth}
\centering
\includegraphics[width=\textwidth]{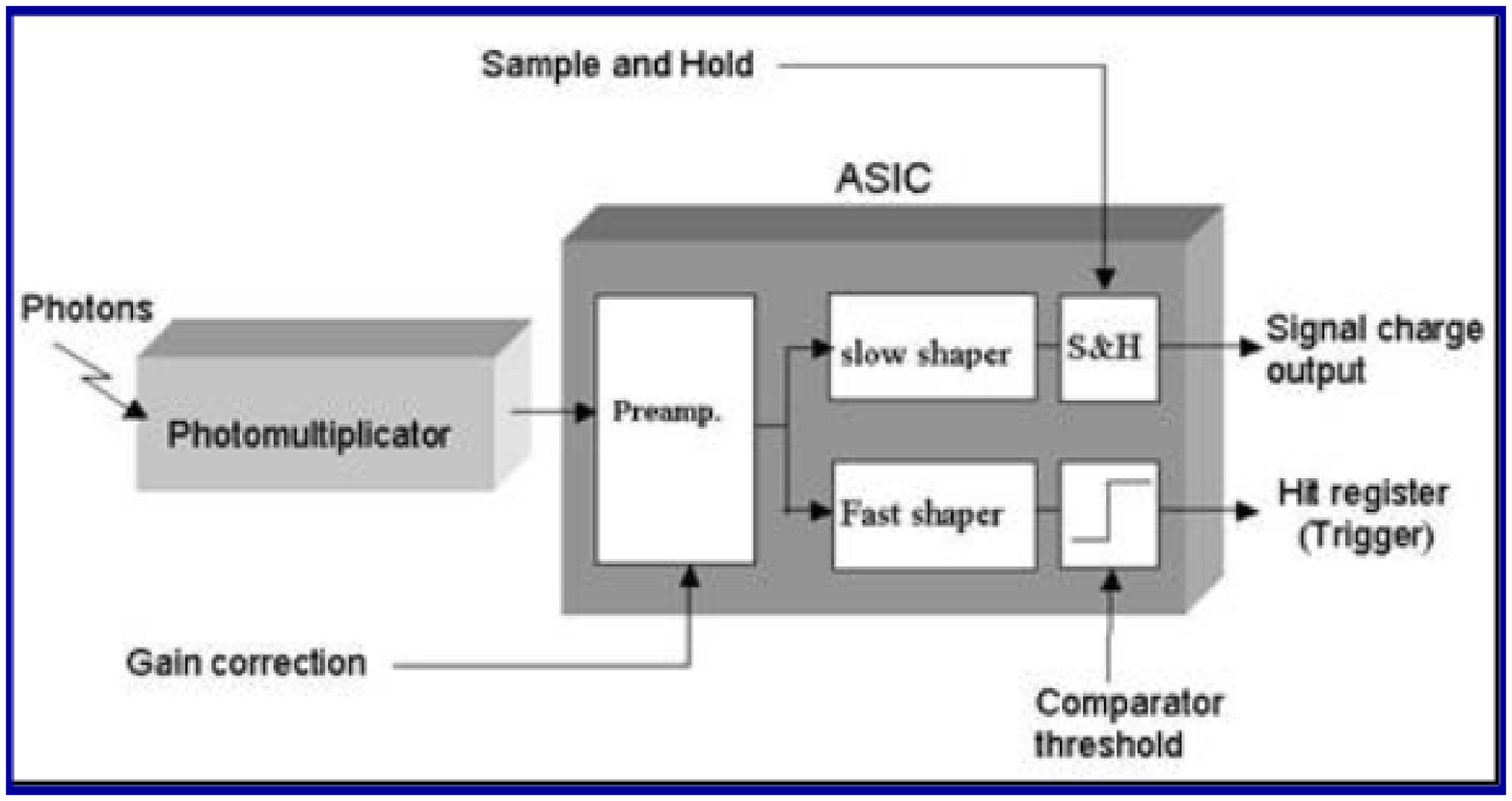}
\caption{\label{fig:OPERAROC}
\it Sketch of the existing Read Out electronics developed for the
OPERA Target Tracker and that is intented to be extended for MEMPHYS
by integrating the ADC and TDC.}	 
\end{minipage}
\end{figure}

The coverage of large areas (around 17,500 m$^2$ for MEMPHYS) with
photodetectors at lowest cost implies a readout integrated electronics
circuit (called ASIC). This makes it possible to integrate: high-speed
discriminator on the single photoelectron (pe), the
digitisation of the charge on 12 bits ADC to provide numerical signals
on a large dynamical range (200~pe), the digitisation of time on 12
bits TDC to provide time information with a precision of 1~ns, and
channel-to-channel gain adjustment to homogenize the response of the
photomultipliers and to thus use a common high voltage. Such an ASIC
for readout electronics allows moreover a strong reduction of the
costs, as well as external components (high-voltage units, cables of
great quality...) since the electronics and the High Voltage may be
put as close as possible to the PMTs and the generated numerical
signals are directly usable by trigger logical units and 
the data acquisition computers (Fig.~\ref{fig:MEMPHYSPMTS}). 

The main difficulty in associating very fast analog electronics and
digitization on a broad dynamic range does not make it possible yet to
integrate all these functions in only one integrated circuit, but
certain parts were already developed separately as for example in the
OPERA Read Out Channel \cite{Lucotte:2004mi}
(Fig.~\ref{fig:OPERAROC}). The evolution of integrated technologies,
in particular BiCMOS SiGe 0.35$\mu$m, now make it possible to consider
such circuits and has triggered a new campaign of research and
development. 

\newpage

\newpage

\section{Detector Performance}
\label{sec:det}
As mentioned above, we consider a massive water \v{C}erenkov detector
{\`a} la UNO \cite{uno} and review the performances of such a detector for
the main physics fields.

\subsection{Proton decay sensitivity}
For proton decay, no specific simulation for MEMPHYS has been carried out yet.
We therefore rely on the study done by UNO, adapting the results to MEMPHYS
(which has an overall better coverage) when possible.
\subsubsection{$p \rightarrow e^+\pi^0$}

\begin{figure}[htb]
 \begin{minipage}[c]{0.44\textwidth}
\epsfig{figure=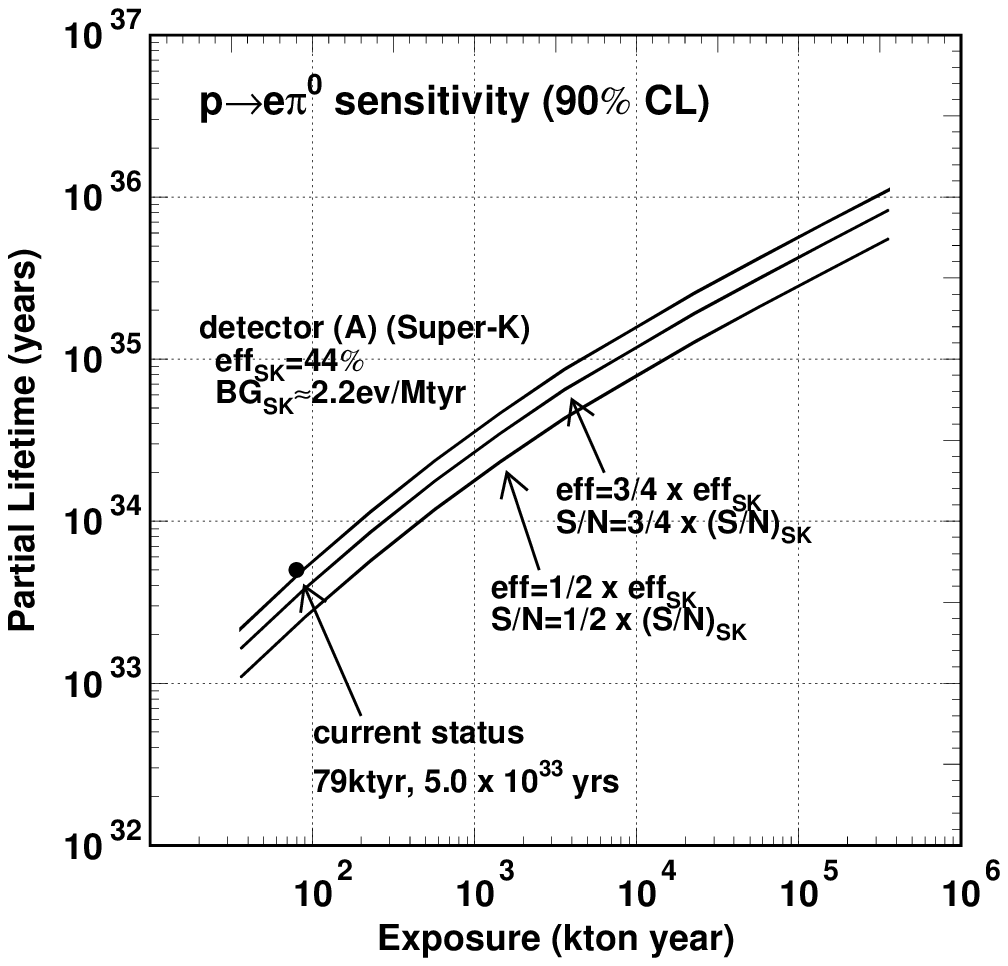,width=\textwidth,angle=0}
\caption{\it \label{pdk1} Sensitivity for $e^+\pi^0$ proton decay
lifetime, as determined by UNO \cite{uno}. MEMPHYS corresponds to case (A).}
\end{minipage}
 \begin{minipage}[c]{0.05\textwidth}
~
\end{minipage}
 \begin{minipage}[c]{0.44\textwidth}
\epsfig{figure=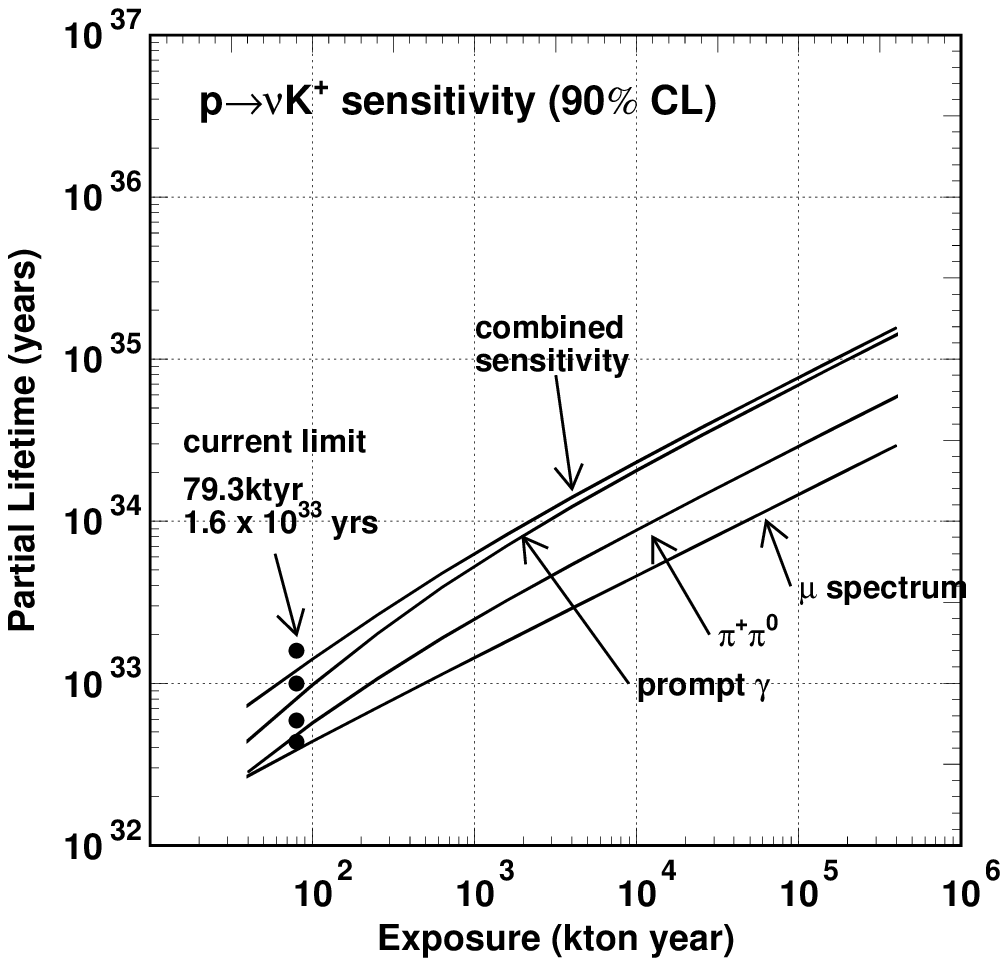,width=\textwidth,angle=0}
\caption{\it \label{pdk9_jbz}
Expected sensitivity on $\nu K^+$ proton decay as a function of MEMPHYS
exposure \cite{uno} (see text for details).}
\end{minipage}
\end{figure}

Following UNO study,
the detection efficiency of $p \rightarrow e^+\pi^0$
(3 showering rings event) is $\epsilon=$43\% 
for a 20 inch-PMT coverage of 40\% or its equivalent, as envisioned for
MEMPHYS. The corresponding estimated
atmospheric neutrino induced background is at the level of 2.25 events/Mt.yr. 
From these efficiencies and background levels,
proton decay sensitivity as a function of detector exposure can be
estimated (see Fig. \ref{pdk1}).
 
$10^{35}$ years partial
lifetime could be reached at the 90\% CL for a 5 Mt.yr exposure with MEMPHYS
(similar to case A in figure~\ref{pdk1}).

\subsubsection{$p \rightarrow \overline{\nu}K^+$}

Since the $K^+$ is below the \v{C}erenkov threshold, this channel is
detected via the decay products of the kaon: a 256 MeV/c muon and
its decay electron (type I) or a 205 MeV/c $\pi^+$ and $\pi^0$
(type II), with the possibility of a delayed (12 ns) coincidence
with the 6 MeV nuclear de-excitation prompt $\gamma$ (Type III).
In Super-Kamiokande, the efficiency for the reconstruction of
$p \rightarrow \overline{\nu}K^+$ is $\epsilon=$ 33\% (I), 6.8\% (II)
and 8.8\% (III),
and the background is at the 2100, 22 and 6/Mt.yr level. For the
prompt $\gamma$ method, the background is dominated by
mis-reconstruction. As stated by UNO, there are good 
reasons to believe that this
background can be lowered at the level of 1/Mt.yr corresponding
to the atmospheric neutrino interaction $\nu p \rightarrow \nu
\Lambda K^+$. In these conditions, and using Super-Kamiokande performances,
a 5 Mt.yr MEMPHYS exposure would
allow to reach the $2\times10^{34}$ years partial lifetime
(see Fig. \ref{pdk9_jbz}).


\subsection{Supernova neutrinos}
\label{sec:SN}

\subsubsection{Core-collapse}
\begin{figure}
\begin{center}
\epsfig{figure=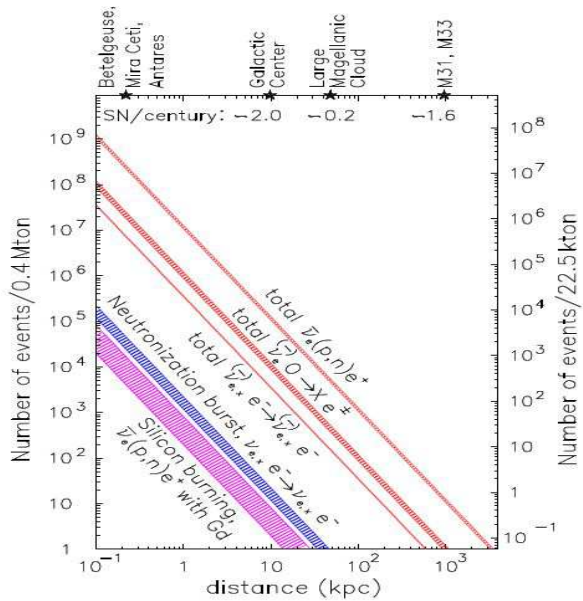,width=8cm,height=8cm}
\caption{\it 
The number of events in a 400 kt water \v{C}erenkov detector (left scale)
and in SK (right scale) in all channels and in the individual
detection channels as a function of distance for a supernova
explosion \cite{Fogli:2004ff}.}
\label{fig:SN}
\end{center}
\end{figure}

The large mass of a MEMPHYS-type detector means that the sample of
events collected during a supernova explosion would outnumber that
of all other existing detectors. For instance, for a supernova at
10 kpc $\sim 2\times 10^5$ events would be observed,
whereas Super-Kamiokande (22.5 kt) will see only 9,000 events (see Figure
\ref{fig:SN}, from ref.~\cite{Fogli:2004ff}). 
These numbers
are to be compared with the 19 (11 for Kamiokande and 8 for IMB)
events coming from the SN1987A in the Large Magellanic Cloud (50 kpc).

An estimated number of $3\pm 1$ supernovae occur in our galaxy and
its satellites every century.
A MEMPHYS-type detector would also be sensitive to supernovae 
occurring throughout
the local group of galaxies. For a supernova explosion in Andromeda 
(730-890 kpc),
the proposed detector will collect roughly the same amount of neutrinos
detected for the SN1987A.
A handful of events might be seen even at a distance as large as 3 Mpc. 

One of the unsolved problems in astrophysics is the mechanism of supernova
core-collapse. 
Inverse beta decay events from the silicon burning phase preceding 
the supernova explosion have very low (sub-threshold) positron 
energies, and could only be detected through neutron capture by adding
Gadolinium \cite{Beacom:2003nk}, 
provided that they can be statistically distinguished from background
fluctuations. 
The silicon burning signal should then be seen with a statistical
significance of 2$\div$8 standard deviations at a reference distance of 1
kpc. Unfortunately, at the 
galactic center ($\sim$10 kpc) the estimated silicon burning signal would
be 100 times smaller and thus unobservable.

There are better prospects to observe the neutronization burst from a
galactic supernova 
by means of elastic scattering on electrons, including contributions
from all flavors: a 0.4 Mton detector might observe such signal with a
statistical significance at the level of 4 standard deviations.
At the distance of the Large Magellanic Cloud, however, the
sensitivity drops dramatically. 

Returning to the overall rate in the inverse beta channel, 
the high statistics available for a 
galactic supernova explosion will allow many possible spectral
analyses, providing insight both on the properties of the collapse
mechanism and on those of neutrinos. 

For the first topic, an example is given in~\cite{Fogli:2004ff} 
in the context of shock-wave 
effects, based on the comparison of arrival times in different energy bins.

Concerning the spectral properties which depend on neutrino
oscillation parameters, 
it has been shown in \cite{Minakata:2001cd} that a detector
like the proposed one, considering the inverse-beta channel alone with
the current best values of solar neutrino oscillation parameters,
would allow the determination of the parameter $\tau_E$, defined as
the ratio of the average energy of time-integrated neutrino spectra
$\tau_E=\langle E_{\bar\nu_\mu}\rangle /\langle E_{\bar\nu_e}\rangle$,
with a precision at the level of few percent, to be compared with a
$\sim$20\% error possible at Super-Kamiokande. This would make it possible to
distinguish normal from inverted mass hierarchy, if
$\sin^2\theta_{13}>10^{-3}$ \cite{Lunardini:2003eh}. 
In the region $\sin^2\theta_{13}\sim (3\cdot 10^{-6}-3\cdot
10^{-4})$, measurements of $\sin^2\theta_{13}$ are possible with a
sensitivity at least an order of magnitude better than planned
terrestrial experiments \cite{Lunardini:2003eh}.

Up to now we have investigate supernova explosions occurring in our galaxy,
however the calculated rate of supernova explosions within a distance of 
10 Mpc is about one per year. Although the number of events from
a single explosion at such large distances would be small, the signal
could be separated from the background with the
request to observe at least two events within a time window 
comparable to the neutrino emission time-scale ($\sim$10 sec),
together with the full energy and time distribution of the 
events \cite{Ando:2005ka}.
In a MEMPHYS-type detector,
with at least two neutrinos observed, a supernova could be identified 
without optical confirmation, so that the start of the light curve 
could be forecasted by a few hours, along with a short list of probable
host galaxies. This would also allow the detection of supernovae
which are either heavily obscured by dust  or are optically
dark due to prompt black hole formation.
Neutrino detection with a time coincidence could therefore act
as a precise time trigger for other supernova detectors 
(gravitational antennas or neutrino telescopes).

Finally, one can notice that electron elastic scattering events would
provide a pointing accuracy on the supernova 
explosion of about $1^\circ$. 

\subsubsection{Diffuse Supernova Neutrinos}

An upper limit on the flux of
neutrinos coming from all past core-collapse supernovae 
(the Diffuse Supernova Neutrinos~\footnote{We 
prefer to denote these neutrinos as ``Diffuse'' rahter than ``Relic''
to avoid confusion with the primordial neutrinos produced one second
after the Big Bang.}, DSN) has been set by the
Super-Kamiokande experiment \cite{Malek:2002ns},
however most of the estimates are below this limit and therefore 
DSN detection thorough inverse
beta decay  appears to be feasible at a megaton scale water \v{C}erenkov
detector.

Typical estimates for DSN fluxes 
(see for example \cite{Ando:2004sb}) predict an event
rate of the order of 0.1$\div$0.5 cm$^{-2}$s$^{-1}$MeV$^{-1}$ 
for energies above 20 MeV, a cut 
imposed by the rejection of spallation events. 
After experimental selections analogous to the ones
applied in the Super-Kamiokande analysis, such events are retained with an
efficiency of about 47\% for energies between 20 and 35 MeV; this is
to be considered as a very conservative estimate at MEMPHYS, where the
bigger overburden will reduce the cosmic-muon induced background and
less stringent selection criteria can be applied. 
Two irreducible backgrounds remain: atmospheric $\nu_e$ and $\bar\nu_e$, 
and decay electrons from
the so called ``invisible muons'' generated by CC interaction of
atmospheric neutrinos and having an energy below threshold for
\v{C}erenkov signal. 

\begin{figure}
\begin{center}
\epsfig{figure=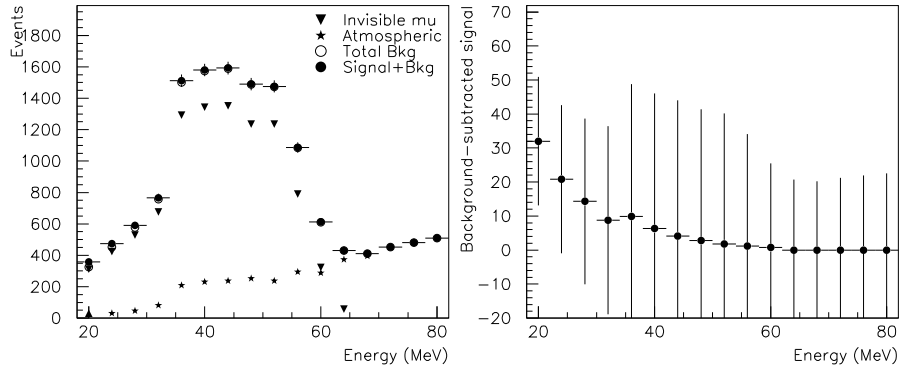,width=13cm}
\caption{\it Diffuse Supernova Neutrino signal and backgrounds (left)
and subtracted signal with statistical errors (right) in a 440 kt
water \v{C}erenkov detector with a 10 years exposure. 
The selection efficiencies of SK were assumed; 
the efficiency change at 34 MeV is due to the spallation cut.}
\label{fig:snr}
\end{center}
\end{figure}

The spectra of the two backgrounds were taken from the 
Super-Kamiokande estimates
and rescaled to a fiducial mass of 440 kton of water, while the
expected signal was computed according to the model called LL
in \cite{Ando:2004sb}. 
The results are shown in Fig.~\ref{fig:snr}:
the signal could be observed with a statistical significance of about
2 standard deviations after 10 years.

\begin{figure}
\begin{center}
\epsfig{figure=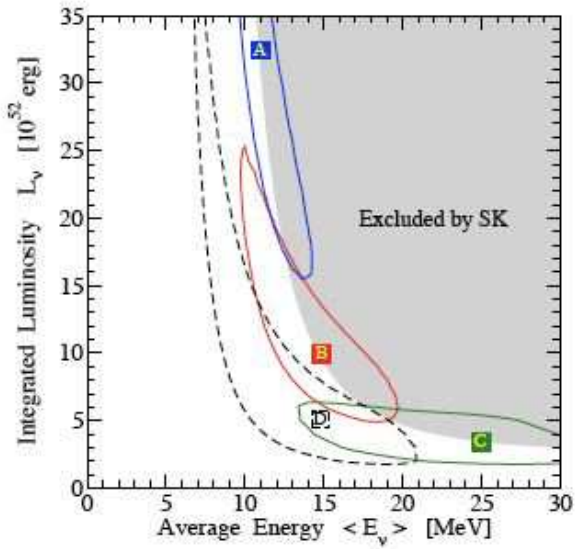,width=8cm}
\caption{\it Possible 90\% C.L. measurement of the emission parameters
of supranova $\bar\nu_e$ emission after 5 years running of
a Gd-enhanced Super-Kamiokande detector, which would correspond to 1 year 
of one MEMPHYS shaft. The points corespond to different assumptions on
the average energy and integrated luminaosty: A,B,C are taken at the
edge of the region excluded by SK, D is often regarded aas the
canonical values for $\bar\nu_e$ emission before neutrino mixing. See
\cite{Yuksel:2005ae}. }
\label{fig:sndpar}
\end{center}
\end{figure}

As pointed out in \cite{Fogli:2004ff}, 
with addition of Gadolinium \cite{Beacom:2003nk} 
the detection of the captured neutron
would give the possibility to reject neutrinos other than $\bar\nu_e$
from spallation events and from atmospheric origin, and the
detection threshold could be lowered significantly - to about 10 MeV -
with a large gain on signal statistics. 
The tails of reactor neutrino spectra would
become the most relevant source of uncertainty on the background. In
such condition, not only would the statistical significance of the
signal become much higher, but is would even be possible to
distinguish between different theoretical predictions. For example,
the three models considered in \cite{Ando:2004sb} 
would give 409, 303 and 172 events respectively above 10 MeV. 
An analysis of the expected DSN spectrum that would be observed
with a Gadolinium-loaded water \v{C}erenkov detector has been carried out
in \cite{Yuksel:2005ae}: the possible limits on the emission parameters of
supernova $\bar\nu_e$ emission have been computed for 5 years running of
a Gd-enhanced SuperKamiokande detector, which would correspond to 1 year 
of one MEMPHYS shaft, and are shown in Fig.~\ref{fig:sndpar}. 
Detailed studies on characterization of the backgrounds, however,
are needed.

\subsection{Neutrino oscillation physics}
\label{sec:oscillations}

\subsubsection{With the CERN-SPL SuperBeam}
\label{sec:CERN-SPL}
 In  the initial CERN-SPL SuperBeam project  
\cite{SPL,SPL-Physics,SPL-Physics2,SPL-Physics3,YELLOW}
 the planned 4MW SPL (Superconducting Proton Linac)  would deliver a 2.2  GeV/c
 proton beam sent on a Hg target to generate
 an intense $\pi^+$ ($\pi^-$) beam focused by a suitable
 magnetic horn in a short decay tunnel. As a result, an intense
 $\nu_{\mu}$ beam is produced
 mainly via the $\pi$-decay,  $\pi^+ \rightarrow \nu_{\mu} \; \mu^+$ providing a
 flux $\phi \sim 3.6 {\cdot} 10^{11} \nu_{\mu}$/year/m$^2$  at 130 Km
 of distance, and an average energy of 0.27 GeV.
 The $\nu_e$ contamination from $K$ is suppressed by threshold effects and
amounts to 0.4\%.
 The use of a near and far detector (the latter 130~km away
 at Fr\'ejus \cite{Mosca}, see Sec.~\ref{sec:undlab})
 will allow for both $\nu_{\mu}$-disappearance and
 $\nu_{\mu} \rightarrow \nu_e$ appearance studies.
 The physics potential of the 2.2 GeV SPL SuperBeam (SPL-SB)
 with a water \v{C}erenkov far detector 
with a fiducial mass of 440 kton,  has been extensively
 studied \cite{SPL-Physics}. 

 New developments show that the potential of the SPL-SB potential could be
 improved by rising the SPL energy to 3.5 GeV \cite{Campagne:2004wt},
 to produce more copious secondary mesons
 and to focus them more efficiently. This increase in energy is made possible
 by using state of the art RF cavities instead of the previously
 foreseen LEP cavities \cite{Garoby-SPL}.

The focusing system (magnetic horns) originally optimized in the context of a 
Neutrino Factory \cite{SIMONE1,DONEGA} has been 
redesigned considering the specific
 requirements of a Super Beam.
 The most important points are that
the phase spaces that are covered by the two types
 of horns are different, and that for a Super Beam the pions to be focused 
should have
 an energy of the order of 800~MeV
to get a mean neutrino energy of $300$~MeV.
The increase in kaon production rate, giving higher \nue contamination,
has been taken into account, and should be refined using HARP results
 \cite{Harp}.

 In this upgraded configuration, the neutrino flux is increased 
by a factor $\sim 3$ 
 with respect to the 2.2 GeV configuration, 
and the number of expected $\nu_\mu$ charged currents 
is about $95$ per ${\rm kton \cdot yr}$ in MEMPHYS.

A sensitivity $\sin^2(2\thetaot) < 0.8 \cdot 10^{-3}$ 
is obtained  in a 2 years $\nu_\mu$ plus
8 year \nubarmu\ run (for $\delta = 0$,
intrinsic degeneracy accounted for, sign and octant
 degeneracies not accounted for), allowing
for a discovery of CP violation (at 3 $\sigma$ level) for 
$\delta \geq 60^\circ$
for $\sin^2(2\thetaot) = 1.8 \cdot 10^{-3}$ 
and improving to $\delta \geq 20^\circ$ for
$\sin^2(2\thetaot) \geq 2 \cdot 10^{-2}$ 
\cite{MMNufact04, Campagne}. These 
 performances are shown in Fig.~\ref{fig:th13}, they are found equivalent to
Hyper-Kamiokande. 
These limits have been obtained first using realistic simulations
based on Super-Kamiokande performances (Background level, signal efficiencies,
and associated systematics at the level of 2\%), and more recently confirmed
using GLoBES \cite{Globes}.

Let us conclude this section by mentioning that further studies of the
SPL superbeam will take place inside the Technical Design Study to be submitted 
to Europe by the neutrino factory community towards the end of 2006.

\subsubsection{With the CERN BetaBeams}
\label{sec:BetaBeam}

 \begin{figure}
  \centerline{\epsfig{file=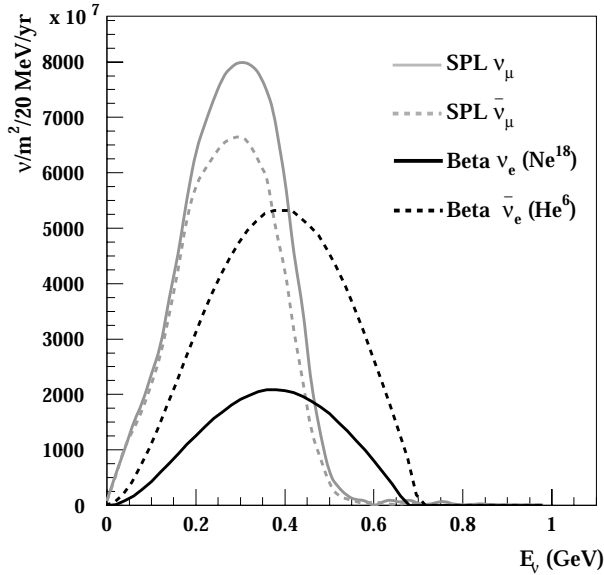,width=0.5\textwidth}}
  \caption{\it Neutrino flux of $\beta$-Beam ($\gamma=100$)
   and CERN-SPL SuperBeam, 3.5 GeV, at 130 Km of distance.}
  \label{fig:fluxes}
 \end{figure}

\begin{figure}
 \centerline{\epsfig{file=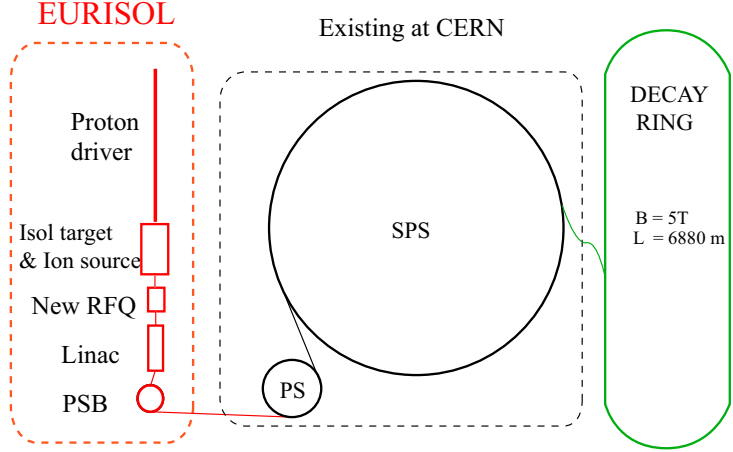,width=0.60\textwidth}  }
\caption{\it 
A schematic layout of the BetaBeam complex. On the left, the low energy part is
largely similar to the EURISOL project \cite{Eurisol}.
 The central part (PS and SPS) uses
existing facilities. On the right, the decay ring has to be built.}
\label{fig:sketch}
\end{figure}

BetaBeams  have been proposed by 
P. Zucchelli in 2001 \cite{Zucchelli:2002sa}.
The idea is to generate pure, well collimated and intense
\nue\  (\nubare) beams by producing, collecting, accelerating radioactive ions
and storing them in a decay ring in 10 ns long bunches, to suppress
the atmospheric neutrino backgrounds.
The resulting BetaBeam  spectra
can be easily computed knowing the beta decay spectrum of the parent
ion and the Lorentz boost factor $\gamma$, and these beams are virtually 
background free from other flavors.
 The best ion candidates so far
 are  $^{18}$Ne  and $^6$He; for \nue\ and \nubare\  respectively.
The schematic layout of a Beta Beam is shown in figure~\ref{fig:sketch}.
It consists of three parts\,:
\begin{enumerate}
\item A low energy part, where a small fraction (lower than 10\%) of the
protons accelerated by the SPL are shot on specific target to produce
$^{18}$Ne or $^6$He; these ions are then collected by an ECR source
of new generation \cite{Sortais} which delivers ion bunches with 100 keV
energy, then accelerated in a LINAC up to 100 MeV/u. This part could be
shared with nuclear physicists involved in 
the EURISOL project \cite{Eurisol,Rubbia:2006pi}.
\item The acceleration to the final energy uses a rapid cycling cyclotron
(labelled PSB) which further accelerates and bunches the ions before sending
them to the PS and the SPS, where they reach their final energy ($\gamma$
around 100). In this process, 16 bunches (150 ns long) in the booster
are transformed into 4 bunches (10 ns long) in the SPS.
\item Ions of the required energy are then stored in a decay ring, with 2500~m
long straight sections for a total length of 7000~m, so that 36\% of the decays
give a strongly collimated and ultra pure neutrino beam aimed at the Fr\'ejus
detector.
\end{enumerate}
A baseline study for the betabeam has been initiated at CERN, and is now
going on within the european FP6 design study for EURISOL.
A specific task is devoted to the study of the high energy part (last 2 items
above). A complete conceptual design for the decay ring has already been
performed. The injection in the ring uses the asymetric merging scheme,
validated by experimental tests at CERN. The actual performances of the new
ECR sources will also be studied with prototypes in the framework of the
EURISOL design study.

The potential of such betabeams sent to MEMPHYS has been studied in the
context of the baseline scenario, using reference fluxes of $5.8 {\cdot}
10^{18}$ \He\ useful
decays/year and $2.2{\cdot}10^{18}$ \Ne\  decays/year, corresponding to a
reasonable estimate by experts in the field of the ultimately
achievable fluxes.

\begin{figure}
    \epsfig{file=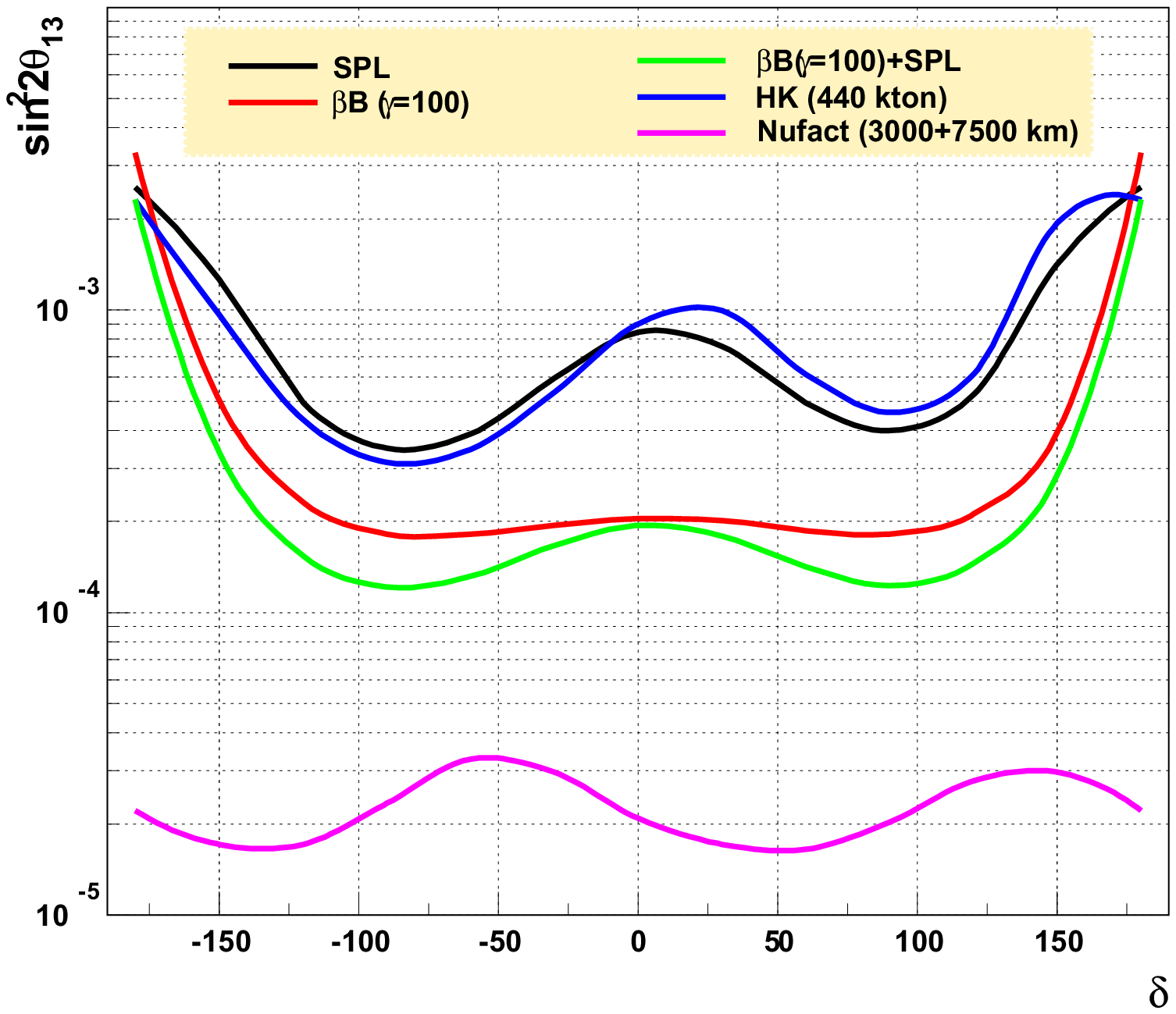,width=0.54\textwidth}   
    \hfill
    \epsfig{file=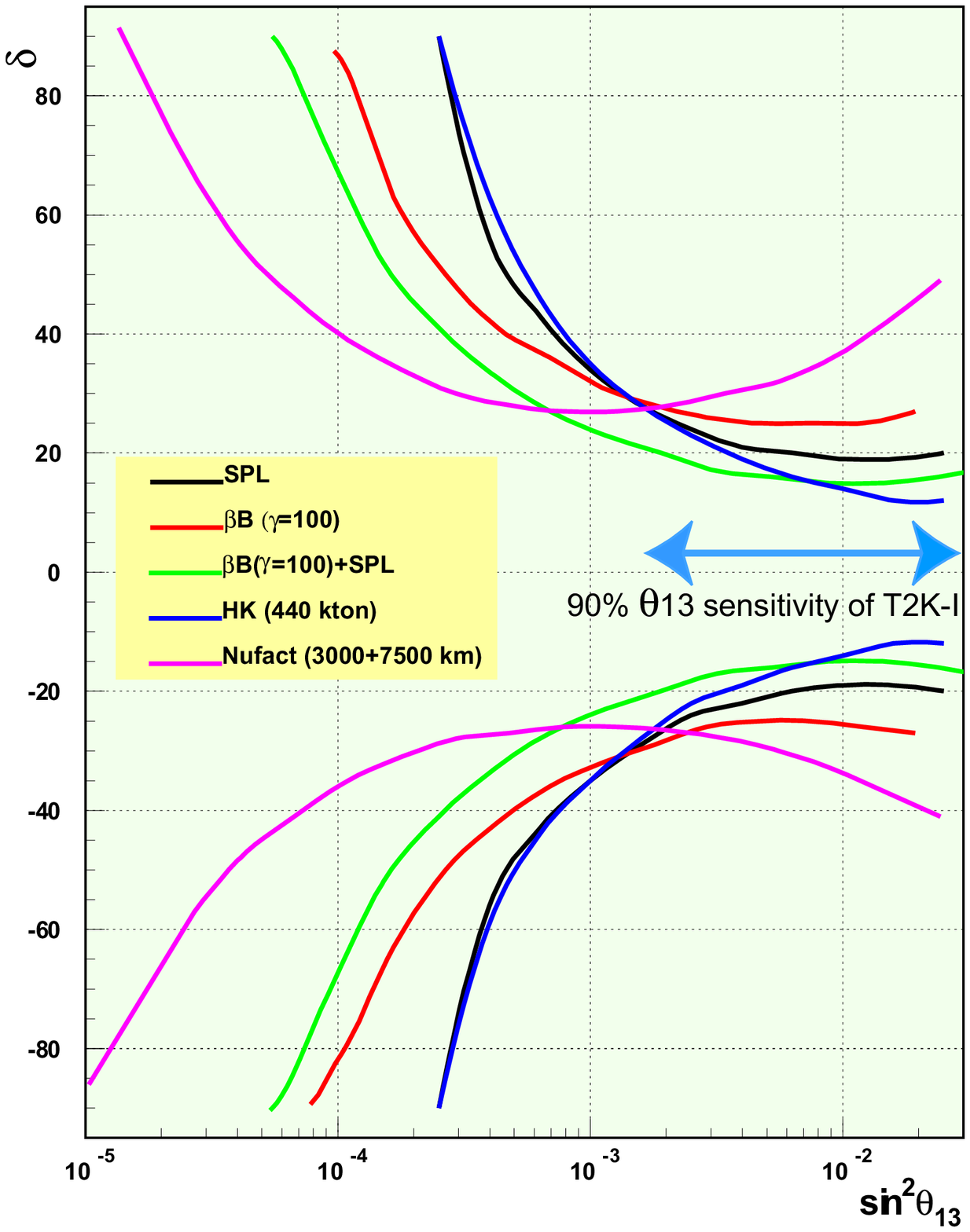,width=0.43\textwidth}  
    \caption{\it LEFT: \thetaot \  90\% 
             C.L. sensitivity as function of $\delta_{CP}$ for
             $\dmtt=2.5{\cdot}10^{-3}eV^2$, $\sigdm=1$, 2\%
             systematic errors.
             SPL-SB  sensitivities have been computed for a
             2 year \numu + 8 year \nubarmu run, $\beta$B ($\gamma$ = 100)
             for a 5 year \nue + 5 year \nubare run, 200 MeV energy bins for
             both beams.
             The combination of SPL-SB and $\beta$B is also shown.
             HK and NuFACT curves are adapted from \cite{VolutaDaAndrea}:
             HK curves corresponds to Hyper-Kamiokande with the same fiducial mass,
             running time and systematics as MEMPHYS, using the 4MW beam from
             JAERI.
             The NuFACT curve corresponds to 5 year runs for each polarity,
             two 50kton iron detectors located at 3000 and 7000 km receiving
             neutrinos from 10$^{21}$ useful 50 GeV muon decays per year,
             detector systematics set at 2\%, matter profile uncertainty set at
             5\%, energy threshold set at 4 GeV.
      RIGHT: $\delta_{CP}$ discovery potential 
             at $3 \sigma$  computed for the same
             conditions.}
  \label{fig:th13}
\end{figure}

First oscillation physics studies 
\cite{Mezzetto:2003ub,Bouchez:2003fy,Mezzetto:2004gs,Donini:2004hu}
 used $\gamma_{\He}=60$ and $\gamma_{\Ne}=100$.
But it was soon realized that the optimal values were actually $\gamma = 100$
for both species, and the corresponding performances are shown in 
figure~\ref{fig:th13}, exhibiting a strong improvement over SPL superbeam
performances, extending the range of sensitivity for 
$\sin^2(2\theta_{13})$ down to $2\cdot 10^{-4}$
and improving CP violation sensitivity at lower values
of $\theta_{13}$.

To conclude this section, let us mention a very recent development 
of the Beta Beam concept leading to the
possibility to have monochromatic, single flavor neutrino beams
by using ions decaying 
through the electron capture process \cite{Bernabeu,Sato}.
A suitable ion candidate exists\,: $^{150}$Dy, whose performances have
been already delineated \cite{Bernabeu}. Such beams would in
particular be perfect to
precisely measure neutrino cross sections in a near detector with the
possibility of an energy scan by varying
the $\gamma$ value of the ions.

For a review of the different  Beta Beam configurations, see~\cite{Volpe:2006in}.

\subsubsection{Combining SPL Super Beam and Beta Beam}
Since betabeams use only a small fraction of the protons available from the
SPL, both beta beam and superbeam can be run at the same time.
The combination of superbeam and betabeam results further improves the
sensitivity on $\theta_{13}$ and $\delta$, as shown on figure~\ref{fig:th13}.
It is better in all cases than Hyper-Kamiokande sensitivity, except maybe for very 
large values of $\sin^2(2\theta_{13})$ above $0.04$ 
The sensitivity on CP violation is even better than that of a neutrino factory
for $\sin^2(2\theta_{13})$ above $3.5\cdot 10^{-3}$ 
(but neutrino factories are still a factor 3
better for $\theta_{13}$ sensitivity).
This combination of super and betabeams offers other advantages, since the
same parameters $\theta_{13}$ and $\delta_{CP}$ may be measured in many
different ways, using 2 pairs of CP related channels, 2 pairs of T related
channels, and 2 pairs of CPT related channels which should all give
coherent results. In this way the estimates of the systematic errors,
different for each beam, will be experimentally cross-checked.
And, needless to say, the unoscillated data for a given beam will give a large
sample of events corresponding to the small searched-for signal with the
other beam, adding more handles on the understanding of the detector
response.

The MEMPHYS  detector performances  in conjunction with  the SPL
SuperBeam and the $\gamma=100$ Beta Beam have been recently revised in
\cite{Campagne:2006yx}. In this paper are also computed the experimental
capabilities of measuring sign$(\Delta{m}^2_{23}) $ and the $\theta_{23}$
octant by combining atmospheric neutrinos, detected with large
statistics in a megaton scale water \v{C}erenkov detector, with
neutrino beams; as initially pointed out in \cite{latestJJ}. Following
these studies, the MEMPHYS detector could unambiguously measure all
the today unknown neutrino oscillation parameters. It's worth to
stress the fact that the short baseline allows to measure leptonic CP
violation without any subtraction of the fake CP signals induced by
matter effects, still having a sizable sensitivity on the mass
hyerarchy determination thanks to the atmospheric neutrinos. 


\subsubsection{Comparison with other projects}
\label{oscComp}
Before the advent of megaton class detectors receiving neutrino 
from a Super Beam
and/or Beta Beam, several beam experiments (MINOS, OPERA, T2K, NoVA) and
reactor experiments (such as Double-CHOOZ) will have improved our knowledge on
$\theta_{13}$.\\
If $\theta_{13}$ is found by these experiments, it will be "big" 
($\sin^2(2\theta_{13})>0.02$) 
and megaton detectors will be the perfect tool to study CP violation,
with no need for a neutrino factory. If on the contrary, only an upper limit
around $5\cdot 10^{-3}$ to $10^{-2}$ is given on $\sin^2(2\theta_{13})$, 
one might consider an
alternative between a staged strategy, starting with megaton detectors, to
explore $\sin^2(\theta_{13})$ down to $3\cdot 10^{-4}$ 
and start a rich program of non
oscillation physics, eventually followed by a neutrino factory if $\theta_{13}$
is not found; or a more aggressive strategy, aiming directly 
at neutrino factories to explore 
$\sin^2(2\theta_{13})$ down to $10^{-4}$ 
but with
no guarantee of success; in the latter case, the non-oscillation physics 
(proton decay, sypernovae) is lost, but would be replaced by precision
muon physics (which has to be assessed and compared with other projects in this
field).\\
There is no doubt that a neutrino factory has a bigger potential than megaton
detectors for very low values of $\theta_{13}$ (below $5\cdot 10^{-3}$), 
and the only 
competition in that case could come from so-called high energy beta-beams.
An abundant litterature has been published on this subject
(see \cite{latestJJ,HighEnergy,HighEnergy2,HuberBB,SuperSPS,MigNufact05}), 
but most authors have
taken as granted that the neutrino fluxes from betabeams
could be kept the same at
higher energies, which is far from evident \cite{MatsPrivate}
and implies a lot of R\&D on the
required accelerators and storage rings before a useful comparison can be made
with neutrino factories.

Presently, the only pertinent comparison is between the several megaton
projects, namely UNO, Hyperkamiokande and MEMPHYS, or their variants using
liquid argon technology (such as FLARE in the USA, GLACIER in Europe).
In this document, we have shown a comparison between Hyperkamiokande and
MEMPHYS, showing a definite advantage for the latter, due to the betabeam.
However, recent variants of Hyperkamiokande using a second detector in Korea
would have to be considered. UNO, for the time being, refers to a study of a
very long baseline (2500 km) neutrino wide band superbeam produced at
Brookhaven, which gives a disappointing sensitivity on $\theta_{13}$ at the
level of 0.02 
(this is due to the fact that this multiGeV beam leads
to high $\pi^0$ backgrounds in a water \v{C}erenkov detector, as explained
before).\\
Liquid argon detector performances have to be studied, but they will probably
suffer from their lower mass for the lower limit on $\theta_{13}$, while
a better visibility of event topologies would probably help for high values
of $\theta_{13}$, when statistics become important and systematics dominate;
all this has still to be carefully quantified.

Let us mention that a unified way to compare different projects has been made
available to the community , this is the GLoBES package \cite{Globes}.
Figure~\ref{fig:th13} in this document was actually produced using GLoBES, and
some of us are actively pursuing GLoBES-based comparisons in the framework of
the International Scoping Study (ISS), with results expected by mid-2006.     
They will also address the best way to solve problems related to the 
degeneracies on parameter
estimates due to the sign of $\Delta m_{23}^2$, the quadrant ambiguity on
$\theta_{23}$, as well as intrinsic (analytic) ambiguities (In the
present document,
we have supposed $\theta_{23}$ equal to 45$^\circ$, and the absence of matter
effects at low energies make the results insensitive to the mass hierarchy).
But the main point is to feed GLoBES with realistic estimates of the expected
performances of the different projects, in terms of background rejection,
signal
efficiencies and the various related
systematic uncertainties. A coordinated effort
to get realistic numbers for the different projects will be, if successful,
an important achievement of the ISS initiative.

\subsection{Solar neutrinos}

  Water \v{C}herenkov detectors have measured the high energy tail 
of the solar $^{8}$B neutrino flux using electron-neutrino 
elastic scattering \cite{Smy:2002rz}. 
Since such detectors could record the time of an interaction and reconstruct 
the energy and direction of the recoiling electron, unique information 
of the spectrum and time variation of the solar neutrino flux was extracted. 
This provided further insights into the ``solar neutrino problem'', 
the deficit of the neutrino flux (measured by several experiments) 
with respect to the flux expected by the standard solar models. 
It also constrained the neutrino flavor oscillation solutions in a fairly 
model-independent way.

The recoiling electrons from solar neutrino interactions are low in energy 
and produce few \v{C}herenkov photons. However, if at least 20 \% of the 
detection surface is photo-sensitive then solar neutrinos above 10 MeV 
could be detected even with a modest photo-sensor efficiency. 
A detector with larger size than any existing  Water \v{C}erenkov  has 
the potential to measure spectrum and time-variation of the high-energy 
solar neutrino flux more precisely, if systematic uncertainties can be 
kept small.
 For example, Super-Kamiokande's measurements obtained from 1258 days 
of data could be repeated in about half a year (the seasonal flux variation 
measurement requires of course a full year). In particular, a first 
measurement of the flux of the rare hep neutrinos may be possible.
Elastic neutrino-electron scattering is strongly forward peaked. 
To separate the solar neutrino signal from background events, this 
directional correlation is exploited. Angular resolution is limited 
by multiple scattering.  The reconstruction algorithm first reconstructs 
the vertex from the PMT times and then the direction assuming a single 
Cherenkov cone originating from the reconstructed vertex. 
Reconstructing 7 MeV events in a 400 kton 
fiducial volume water \v{C}erenkov (UNO,MEMPHYS,...) seems not to be a problem.

This means we are able to make improvements in solar neutrino detection 
with a megaton-scale \v{C}erenkov
detector: even if it is not the main goal of such a detector 
it could be an excellent by-product.

\newpage
\section{Schedule}

The following table presents an optimal schedule for the 
European project taking into account the key date of the completion 
of the new tunnel excavation around 2010. Soon after, CERN will have to decide
its post-LHC strategy, while nuclear physicists will hopefully choose CERN as
the host laboratory for the EURISOL project. 
We would also like to stress that
the schedule of the neutrino beams from CERN is not constraining the 
start of the other non accelerator items of research. 
\begin{figure}[htb]
\vspace{4cm}
\epsfig{figure=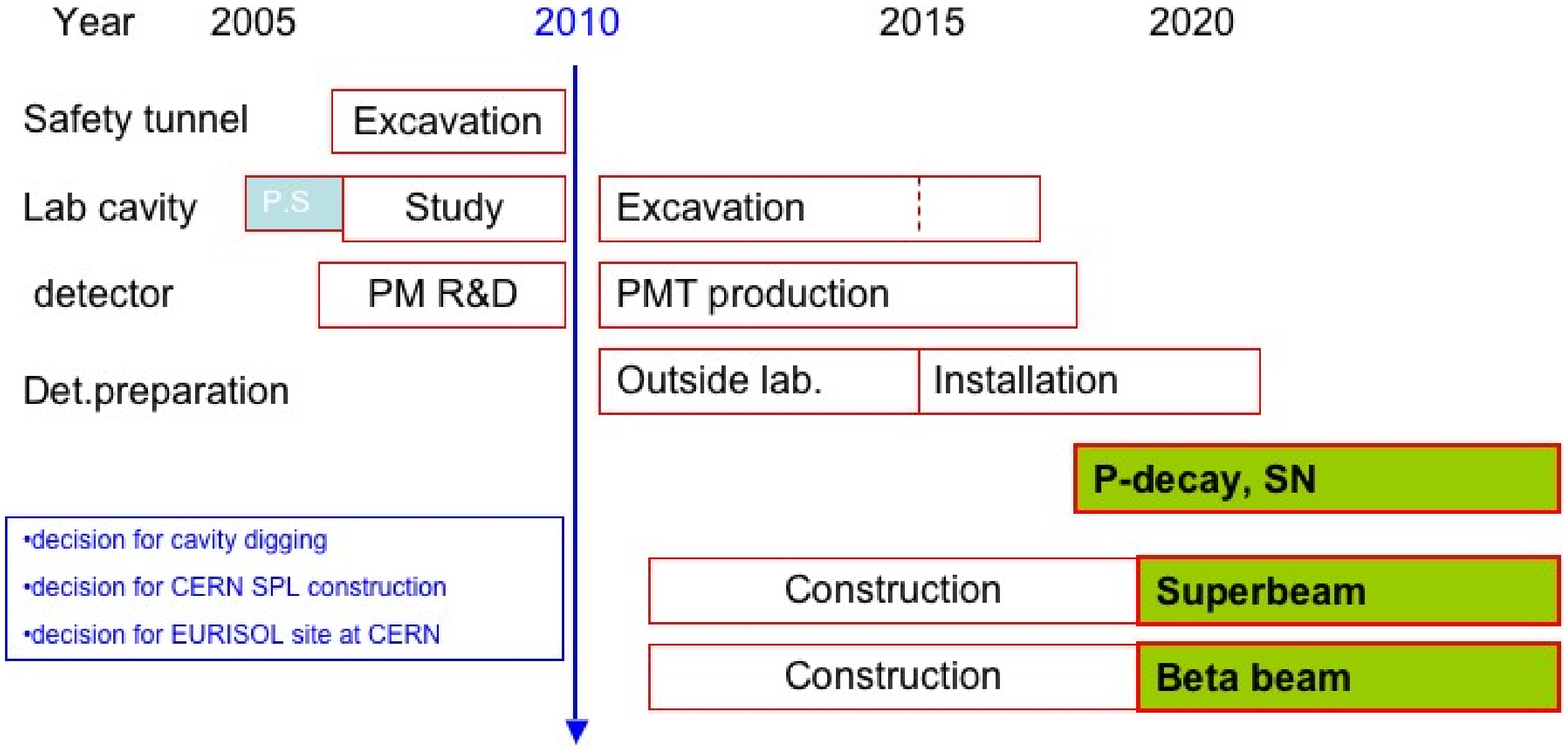,width=\textwidth,angle=0}
\end{figure}

\newpage
\section{Conclusions}
In conclusion a megaton scale 
Water \v{C}erenkov detector at the Frejus site will address a series of 
fundamental issues :
\begin{itemize}
\item explore the nucleon decay with a sensitivity an order of magnitude 
better than current limits  on different channels
\item in the case of a galactic or near galactic supernova explosion, 
track the explosion in unprecedented detail providing at the same time 
information on the third oscillation angle beyond what is currently 
achievable in terrestrial experiments
\item provide a trigger for supernova explosions 
for other astroparticle detectors for supernova exploding in a range of up to
3 Mpc, knowing that 1 supernova explosion per year is expected 
within a distance of 
10 Mpc
\item provide a 4 sigma detection of diffuse supernova
neutrinos after 2-3 years of operation 
\item  in association with a superbeam and betabeam from CERN 
obtain a sensitivity to the third oscillation angle down to   
$\sin^2(2\theta_{13}) \sim 10^{-4}$ and detect maximal CP violation at 3 sigmas  
for $\sin^2(2\theta_{13})$ larger than $3\cdot 10^{-4}$
\end{itemize}
A series of other physics topics, not mentioned here, will also be adressed\,:
for instance neutrino physics, as well as 
interdisciplinary topics in rock mechanics, geobiology, geochemistry, 
geohydrology, geomechanics and geophysics that could benefit
from a large scale underground excavation.

We believe that our project compares favorably with other similar
projects around the world, and should be seriously considered as a
very attractive major European project after the LHC. The proposed
strategy is thus the following: a megaton-scale detector could be
installed at Fr{\'e}jus and start physics in 2018. It would start
proton decay and supernova searches, which would last several
decades. As soon as the neutrino beam from SPL is available,
neutrino oscillation studies can start, and the advent of a beta beam
would increase significantly the performances of the detector.

The signatories are eager to see the MEMPHYS project
come to life. They are aware that the actual location of a megaton detector
will depend on many issues, in particular the share of future big equipments
(such as linear colliders) worldwide. They are prepared to do the proposed
physics in any country, and have already set up collaborations with their
japanese and american colleagues. An inter-regional yearly (US-Europe-Japan) 
workshop series NNN-XX (Next generation of Nucleon decay and Neutrino Physics 
detectors) organizes and structures this convergence of interests. 
The authors of this document hope however that Europe will not
miss a unique opportunity to keep a leading role in the underground physics,
complementary to the Gran Sasso.

Furthermore, it is obvious that the current proposal is complementary 
to other proposals for large undergrounds detectors using 
liquid scintillator (LENA) or liquid argon technologies (GLACIER) 
in order to pursue the same physics goals. 
The advantage of the water \v{C}erenkov technique lies on the possibility 
to instrument very large masses, while liquid argon detectors 
can have an excellent resolution and liquid scintillators 
very low detection thresholds for neutrino physics. 
On the technology side the water \v{C}erenkov seems a straightforward 
extension of the existing techniques while for instance the liquid argon  
option presents daring technological challenges. The realisation of the 
complementarities in physics potential and the common R\&D issues 
(large underground caverns and containers: excavation issues and safety, 
large area low cost photodetection and electronics, 
purification and background issues, interdisciplinary issues, etc.) 
prompted the proponents of the above solutions to start federating 
their efforts in order to  exploit the possible synergies in view of 
common future proposals to the European Union ~\cite{Laguna}
and elsewhere.

\section {Acknowledgements}

The authors would like to thank the engineers of the IN2P3-CNRS laboratories, 
especially Ch. de La Taille (LAL) and J. Pouthas (IPNO), 
for their decisive contributions.

\newpage
\bibliography{Frejus}

\end{document}